\documentclass[twocolumn,prd,aps,superscriptaddress,showpacs,
nofootinbib,preprintnumbers,floats,floatfix]{revtex4}

\usepackage{amsmath}
\usepackage{amssymb}
\usepackage{latexsym}
\usepackage{graphicx,subfigure,graphics,epsfig,amsmath,amssymb}
\usepackage{hyperref}
\usepackage{bm}

\begin{document}

\title{Viable inflationary models ending with a first-order phase
transition}

\author{Marina Cort\^{e}s}
\affiliation{Astronomy Centre, University of Sussex, Brighton BN1
9QH, United Kingdom}
\affiliation{Berkeley Lab, Berkeley, CA 94720, USA}

\author{Andrew R. Liddle}
\affiliation{Astronomy Centre, University of Sussex, Brighton BN1
9QH, United Kingdom}

\begin{abstract}
We investigate the parameter space of two-field inflation models where
inflation terminates via a first-order phase transition causing
nucleation of bubbles. Such models experience a tension from the need
to ensure nearly scale invariant density perturbations, while avoiding
a near scale-invariant bubble size distribution which would conflict
observations.  We perform an exact analysis of the different regimes
of the models, where the energy density of the inflaton field ranges
from being negligible as compared to the vacuum energy to providing
most of the energy for inflation. Despite recent microwave anisotropy
results favouring a spectral index less than one, we find that there
are still viable models that end with bubble production and can match
all available observations. As a by-product of our analysis, we also
provide an up-to-date assessment of the viable parameter space of
Linde's original second-order hybrid model across its full parameter
range.
\end{abstract}

\pacs{98.80.Cq}
\preprint{}

\maketitle
\section{Introduction}\label{intro}

One of the open questions in inflationary cosmology is the mechanism
by which inflation came to an end. The current literature is dominated
by two paradigms, violation of slow roll bringing inflation to an end
while the field is still evolving, and a second-order phase transition
of hybrid inflation type. However, Guth's original (but unsuccessful)
proposal \cite{guth} invoked a first-order phase transition whereby
inflation ended by nucleation of bubbles of true vacuum. First-order
transitions have subsequently experienced bursts of popularity. In the
late 1980s, La and Steinhardt \cite{la_steinhardt} initiated intensive
investigation of `extended inflation' models, where modifications to
Einstein gravity allowed bubble nucleation to complete in single-field
inflation. A few years later those models were struggling in face of
observations, and focus instead returned to Einstein gravity, now in a
two-field context with one rolling and one tunnelling field
\cite{linde, adams_freese, copeland_al}, although see \cite{notari1,notari2}.

In addition to the usual quantum fluctuation mechanism, first-order
inflation models produce density perturbations through the bubble
collisions and subsequent thermalization.  The spectrum of bubble
sizes produced must be far from scale invariance to avoid clear
violation with observed microwave anisotropies --- the largest of the
bubbles would otherwise be blatantly visible
\cite{liddle_wands91,liddle_wands92, griffiths_al}. This requirement
is typically at odds with the need to maintain scale invariance in the
spectrum produced by quantum fluctuations, a tension sufficient to
exclude extended inflation variants except in extremely contrived
circumstances \cite{liddle_wands}. The purpose of this paper is to
investigate whether the strengthened constraints of the post Wilkinson
Microwave Anisotropy Probe (WMAP) era have eliminated the
Einstein gravity first-order models too and, by implication,
assess whether it is plausible that voids exist below current
detection limits.

In Guth's original model, with one field, the inflaton must remain in
the metastable vacuum long enough to allow for sufficient $e$-folds of
inflation but in this case inflation never ends, the bubbles never
thermalize and the transition doesn't complete.  Introduction of a
second field allows a time-dependent nucleation rate, permitting
enough inflation to occur while the nucleation rate is low and a
successful end when the rate rises to high enough values. This idea
was proposed independently by Linde \cite{linde} and, in more detail,
by Adams and Freese \cite{adams_freese} under the name `double-field
inflation'.

Typically the second field, which is trapped in the metastable vacuum,
also provides most of the energy density for inflation, although this
depends on the particular values of parameters chosen. In that regime,
the usual prediction is for a blue spectrum of density perturbations,
$n_{\rm S}>1$.  In the last few years the trend in cosmic microwave
background (CMB) observations has been a tightening of the confidence
limits around a central value $n_{\rm S}$ smaller than one,
disfavouring this regime.  Since our goal is to investigate the
general viability of this type of model we will probe the entire
parameter space, including the intermediate region where the
contributions of each field to the energy density are comparable,
making no approximations based on inflaton or the false vacuum
domination.

As stated above one expects these models to run into difficulty with
recent observations closing in on a nearly scale invariant scalar
spectrum. CMB anisotropies observations place constraints on the
maximum size of bubbles that survive from a first-order phase
transition, at the time when scales of cosmological interest leave the
horizon. In turn this places a strong upper limit on the nucleation
rate at this time, after which it must rise sufficiently to complete
the transition and provide a graceful exit for inflation. In order to
meet these two requirements the field must proceed swiftly along the
potential, what, in light of observations, places the model under
stress.

\begin{figure*}[t]
\begin{center}
$\begin{array}{c c}
    \epsfxsize=8.5cm
    \epsffile{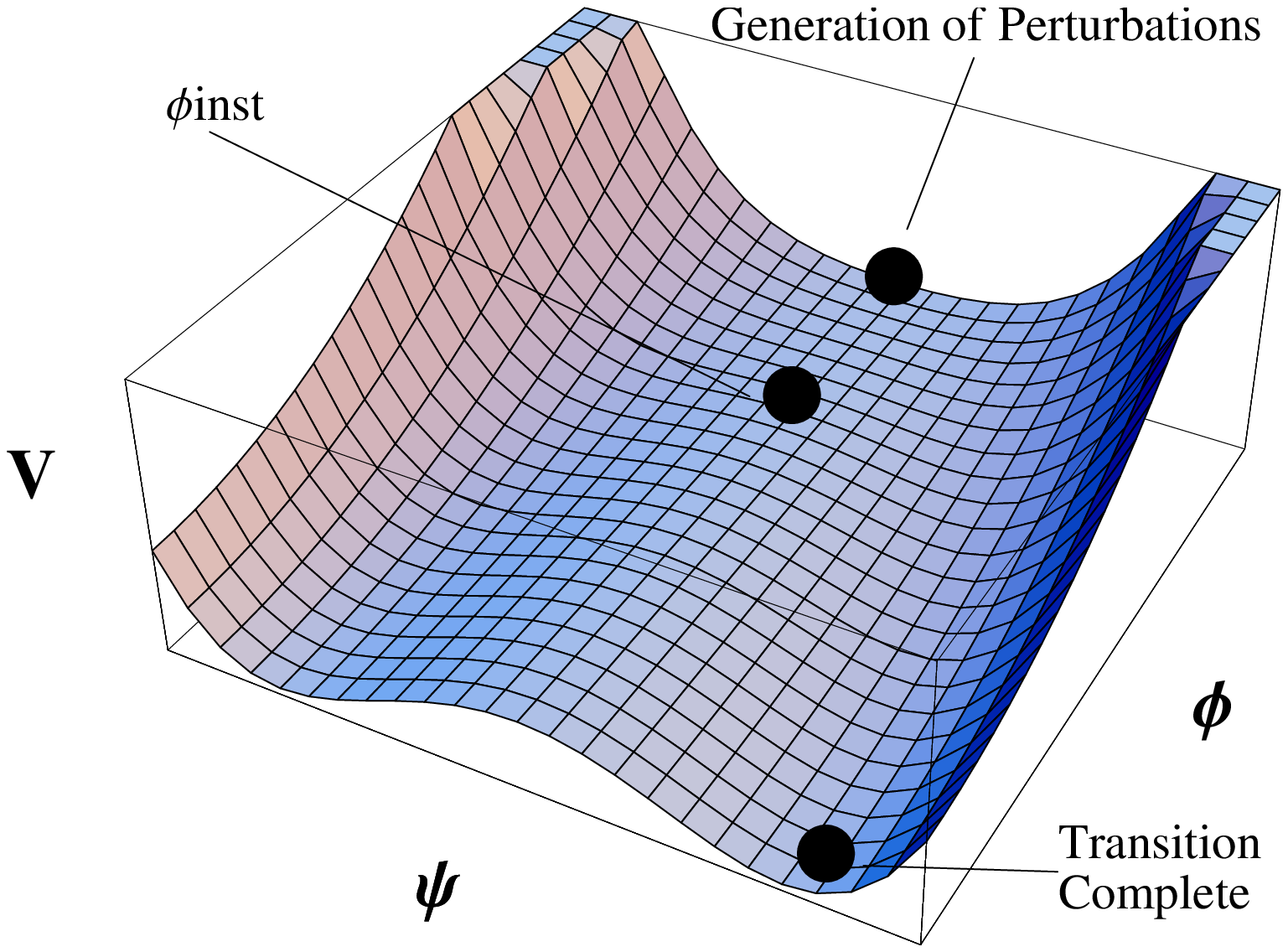} &
    \epsfxsize=8.5cm
	\epsffile{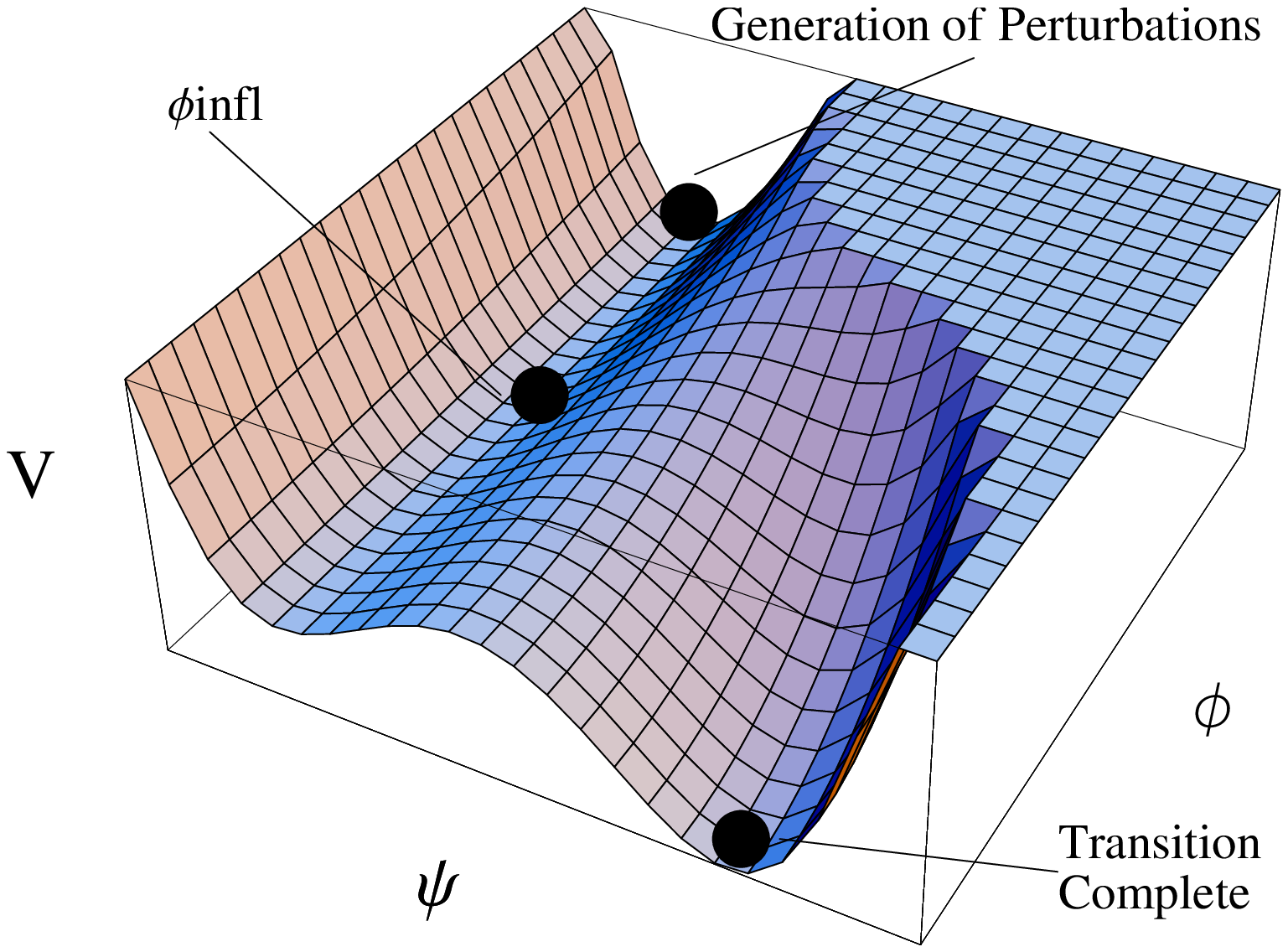}\\
\mbox{\bf (a)} & \mbox{\bf (b)}
\end{array}$
\end{center}
\caption{ \textbf{(a)} The potential for a second-order phase
transition. The field reaches the true vacuum through a continuous
transition, and the breaking of the symmetry implies that there will
be defect formation at the end of the transition.  The true vacuum
minima develop once the field passes the point of instability,
$\phi_{\rm inst}$  \textbf{(b)} The
same for the first-order case.  In this case the
transition is discontinuous and proceeds through quantum tunneling of
the $\psi$ field to the true vacuum. The second minimum develops after
the point of inflection $\phi_{\rm infl}$. The couplings in both
\textbf{(a)} and  \textbf{(b)} have been chosen so as to 
produce a visible barrier height (in working models this is
negligible compared to the false vacuum energy).}
\label{2pot}
\end{figure*}

\section{The first-order model}

We consider throughout a fairly general form of the potential for a
first-order phase transition, given by Copeland et al \cite{copeland_al}.
\begin{eqnarray}\label{pot}
V(\phi,\psi)&=&\frac{1}{4}\lambda(M^4+\psi^4)+ \frac{1}{2}\alpha M^2
\psi^2 -\frac{1}{3}\gamma M \psi^3 \nonumber\\ 
&& + \frac{1}{2} m^2 \phi^2 + \frac{1}{2} \lambda'  \phi^2 \psi^2 \,.
\end{eqnarray}
This extends the simplest second-order hybrid inflation model by
addition of the cubic term for the $\psi$ field. As in conventional
hybrid inflation, one envisages that initially the inflaton field
$\phi$ is displaced far from its minimum, and the auxiliary field
$\psi$ is then held in a false vacuum state by its coupling to the
inflaton.  Perturbations are generated during this initial phase as
$\phi$ rolls slowly along the flat direction. The dynamics in this
region are pretty much those of single-field slow-roll inflation,
though the auxiliary field $\psi$ may provide most of the energy
density for inflation, see Fig.~\ref{2pot}.

In a model where the phase transition is second-order, shown in
Fig.~\ref{2pot}a, the false vacuum becomes unstable after $\phi$
passes a certain value, $\phi_{\rm inst}$, and the fields evolve
classically to their true vacuum (here producing topological defects
as causally separated regions make independent choices as to which
minimum to finish in). Although not the main topic of this paper, we
explore current constraints on this model in the Appendix.

In the first-order case, shown in Fig.~\ref{2pot}b, if the parameters
in Eq.~(\ref{pot}) are chosen appropriately, a second minimum develops
once the field evolves past a point of inflection, $\phi_{\rm
infl}$. At this point bubbles of the true vacuum begin to nucleate and
expand at the speed of light. The percolation rate is initially very
small as the vacuum energies are comparable, but as $\phi$ approaches
zero the interaction between the fields triggers a steep rise in the
bubble production. Inflation ends when the nucleation rate reaches
high enough values that the bubbles percolate and thermalize. In this
case there is only one true vacuum and hence no topological
defects. The channel in which the field rolls after tunnelling is much
too steep to sustain any inflation within the bubbles.

\begin{figure*}[t]
\begin{center}
$\begin{array}{c c c}
    \epsfxsize=5.6cm
    \epsffile{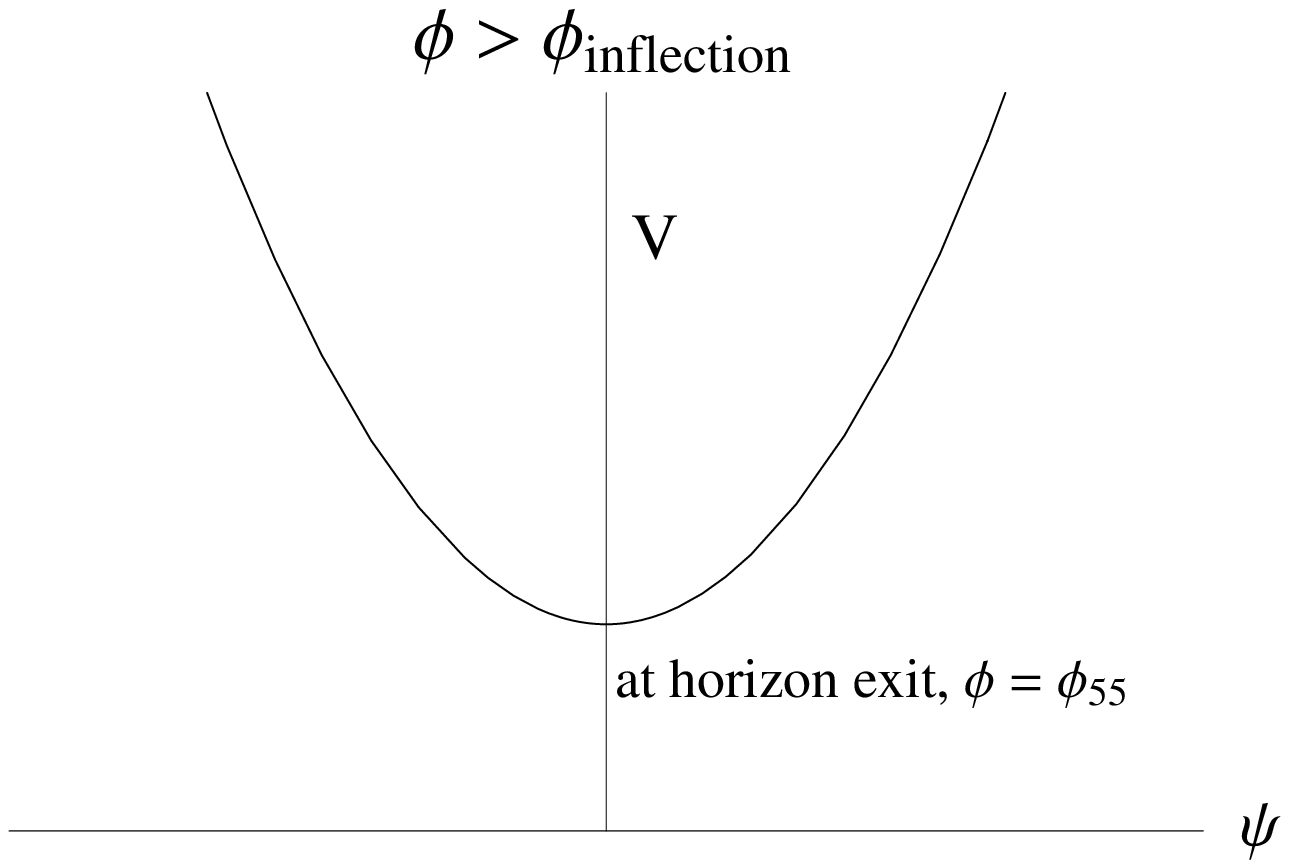} &
    \epsfxsize=5.6cm
	\epsffile{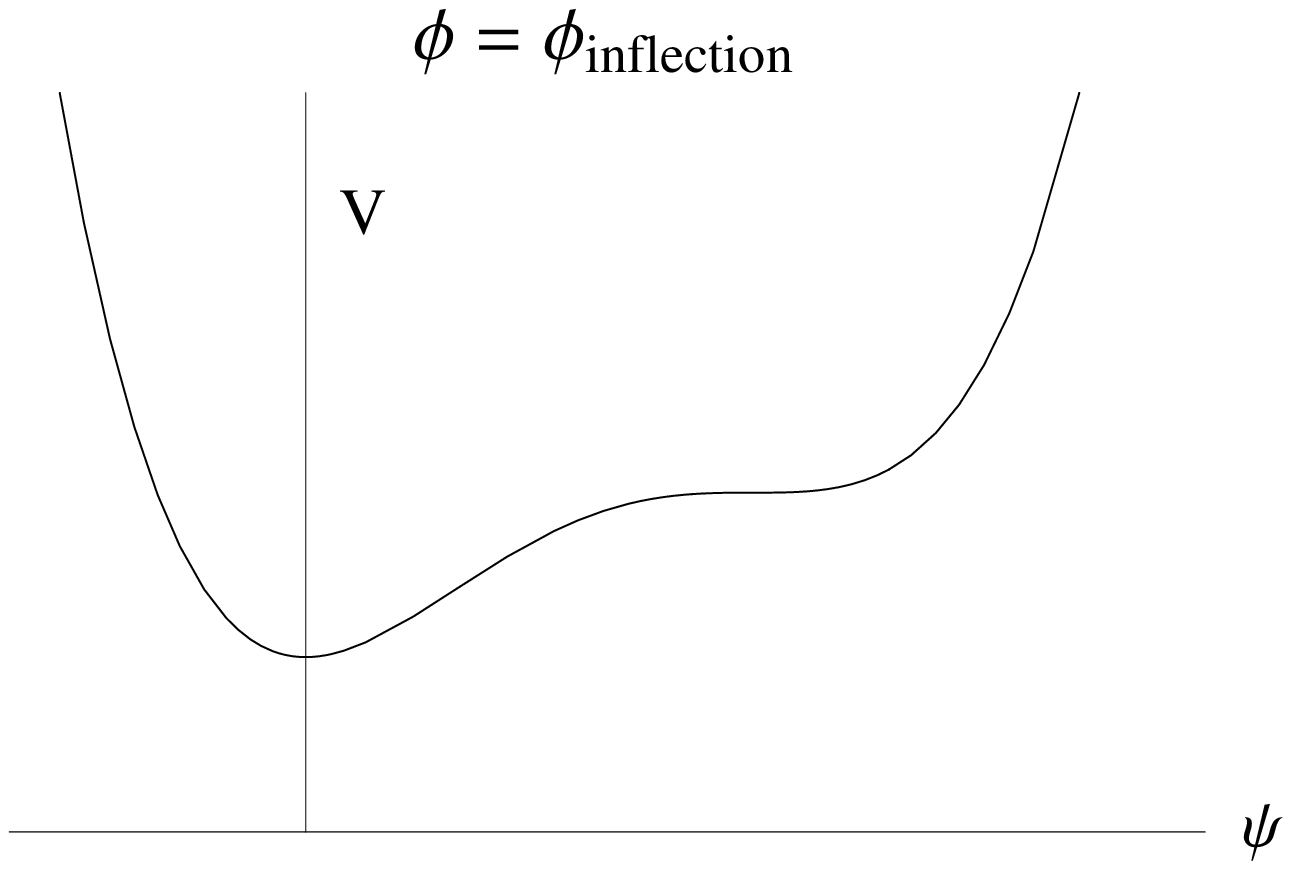} &
	\epsfxsize=5.6cm
	\epsffile{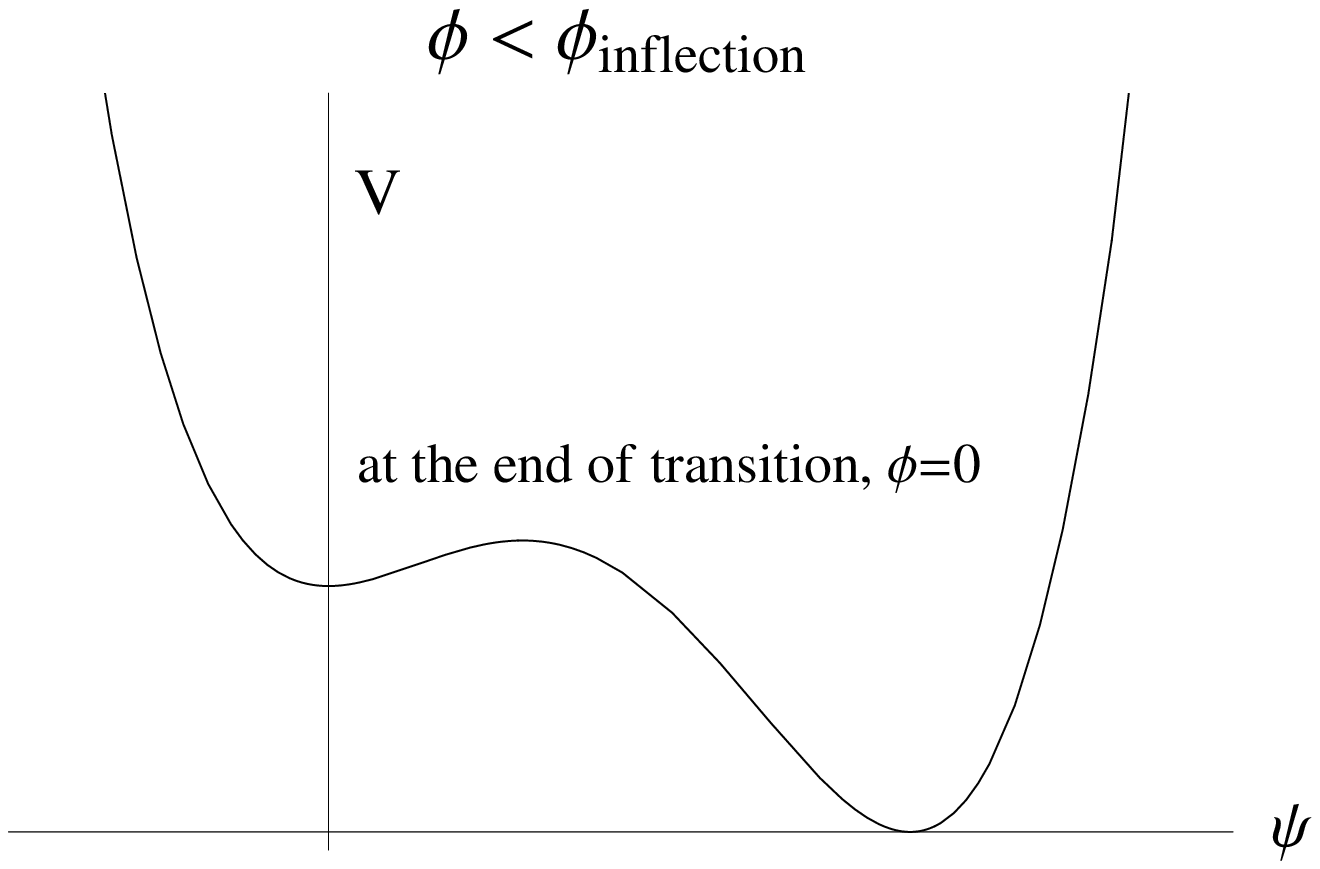} \\
\mbox{\bf (a)} & \mbox{\bf (b)} & \mbox{\bf (c)}
\end{array}$
\end{center}
\caption{ \textbf{(a)} At early times, away from $\phi=0$, there is only
minimum available for $\psi$ and the field is trapped in the false
vacuum.  \textbf{(b)} Given appropriate choices for the couplings a
second minimum begins to develop when the field reaches the point of
inflection of the potential.  \textbf{(c)} Once the transition becomes
energetically favourable the $\psi$ field begins to tunnel to the newly formed minimum, which eventually becomes the true vacuum.}
\label{3reg}
\end{figure*}

For large values of $\phi$ there is only one minimum of the potential,
and in the $\psi$ direction the potential looks like
Fig.~\ref{3reg}a.  However if $\gamma^2 > 4\alpha \lambda$, a second
minimum develops after $\phi$ reaches a point of inflection
\begin{equation}
\phi_{\rm infl}^2=M^2 \frac{\gamma^2-4 \alpha \lambda}{4 \lambda'}\,,
\end{equation} 
as in Fig.~\ref{3reg}b. The presence of the cubic term in the
potential then breaks the degeneracy between the two minima, making it
possible for the field to tunnel to the newly formed minimum. It is
this second minimum that eventually becomes the true vacuum and the
$\psi$ field begins to tunnel once the transition becomes
energetically favourable, Fig.~\ref{3reg}c.

As mentioned in the previous section the quantum generation of
perturbations occurs away from this minimum, while the inflaton is
rolling in the $\phi$ direction, and we consider horizon exit to occur
around 55 $e$-folds before the end of inflation \cite{liddle_leach}. This
evolution of $\phi$ is a crucial feature of the model since it is the
introduction of a time dependence in the tunneling rate that will
allow the phase transition to complete, bringing inflation to an end.

The rate at which the bubbles nucleate is given by the percolation
parameter (the number of bubbles generated per unit time per unit
volume),
\begin{equation}
p=\frac{\Gamma}{H^4}\,.
\end{equation}
In the limit of zero temperature (taken because the transition occurs
during inflation) the nucleation rate of bubbles can be approximated
by \cite{callan_coleman},
\begin{equation}\label{perc}
p=\frac{\lambda M^4}{4H^4}\exp(-S_{\rm E}) \,,
\end{equation}
where $S_{\rm E}$ is the four-dimensional Euclidean action. $S_{\rm
  E}$ was obtained for first-order transition quartic potentials by Adams
  \cite{adams},   who fitted the result as
\begin{equation}
S_{\rm E}= \frac{4 \pi^2}{3 \lambda}(2-\delta)^{-3}(\alpha_1
\delta+\alpha_2 \delta^2+\alpha_3 \delta^3) \,,
\label{4act}
\end{equation}
where $\alpha_1= 13.832 ,~\alpha_2=-10.819,~\alpha_3=2.0765$, and
$\delta$ is a monotonic increasing function of $\phi^2$,
\begin{equation}\label{delta}
\delta=\frac{9 \lambda \alpha}{\gamma^2}+\frac{9 \lambda \lambda'
\phi^2}{\gamma^2 M^2} \,.
\end{equation}
The allowed range has $0<\delta<2$ (outside this range solutions
correspond to energetically disallowed transitions).

The transition to the true vacuum is complete once the percolation
reaches unity, (one bubble per Hubble time per Hubble volume),
allowing the bubbles of the true vacuum to coalesce.

However in the most general case inflation need not end through bubble
nucleation.  If the potential is too steep slow-roll is violated
before bubbles thermalize and inflation ends before the transition
completes. In this case the precise mechanism which completes the
transition is irrelevant given that it occurs after inflation ends,
and for our purposes the scenario is indistinguishable from the
single-field case. (In this paper we do not consider gravitational
waves produced via bubble collisions, but these may provide a further
observable \cite{hogan, kosowsky_al, huber_konstandin, caprini_al}
that can ultimately be used to constrain this type of model.)

The distinction between the two possibilities is given by the two
values of the field, that at which the nucleation rate reaches unity,
and that which makes \mbox{$\epsilon \sim 1$} (violation of
slow-roll), where $\epsilon$ is the usual slow roll parameter defined
in Eq.~(\ref{eps}). Inflation ends by whichever value of $\phi$ is
reached first,
\begin{equation}
\phi_{\rm end}=\max(\phi_{\rm \epsilon},\phi_{\rm crit}) \,.
\end{equation}

\section{Inflationary dynamics}

\subsection{Regimes}

Two different regimes can be distinguished, regarding which field we
wish to have dominate the energy density. In the usual hybrid
inflation regime the energy density of the potential is dominated by
the false vacuum $\lambda M^4 \gg m^2 \phi^2$, which provides the energy
for inflation. In the opposite regime, in which the inflaton dominates
the energy density, the dynamics rapidly approach those of
single-field inflation since, as we will see, slow-roll violation
occurs sooner.

Working in either of these two regimes would allow us to simplify some
of the expressions governing the dynamics during inflation, such as
the number of $e$-folds and the slow-roll parameters, Eqs.~(\ref{Ne}),
(\ref{eps}) and (\ref{eta}), and to proceed via an analytical
treatment instead of a numerical one.  However our purpose here is to
probe the dynamics of the full $n_{\rm S}-r$ parameter space, ($r$ is
the tensor-to-scalar ratio given by Eq.~(\ref{r})) so as to determine
whether there still remain models consistent with CMB
observations. Hence we also include the intermediate regime in our
analysis, where the energy densities of the two fields are comparable,
particularly when the transition between slow roll violation and
bubble nucleation occurs. For this reason we will retain the full form
of the potential and proceed through numerical calculations.

\subsection{Field dynamics}

In order to specify the dynamics of each model we begin by finding the
field value, $\phi_{\rm max}$, at which inflation ends so we need to
determine $\phi_{\rm \epsilon}$ and $\phi_{\rm crit}$. $\phi_{\rm \epsilon}$ is obtained by evaluating the first slow-roll
parameter for our potential and taking it to unity, \footnote{The field $\psi$ sits in the false vacuum during the inflationary phase, since this is the only minimum available to $\psi$ in this region of the potential. This happens regardless of the means to ending inflation, so $\psi$ is set to zero throughout this section. }
\begin{equation}
\epsilon \equiv \frac{m_{\rm Pl}^2}{16 \pi}
\left(\frac{V'}{V}\right)^2 = \frac{m^4 \phi^2 \, m_{\rm
Pl}^2}{\pi (\lambda M^4 + 2 m^2 \phi^2)^2}\approx 1 \,.
\end{equation}
Inverting for $\phi$ yields, 
\begin{equation}\label{eps}
\phi^2_{\epsilon} = \frac{m^2 m_{\rm Pl}^2 \pm m m_{\rm Pl} \sqrt{m^2
    m_{\rm Pl}^2 -8\pi\lambda M^4}-4\pi\lambda M^4}
{8 \pi m^2} \,,
\end{equation}
and we take the largest value of $\phi$. Note that the solution exists
only for large values of $m$, where $m^2 m_{\rm Pl}^2 > 8 \pi \lambda
M^4$.

\begin{figure*}[t]
\includegraphics[width= 0.94 \textwidth]{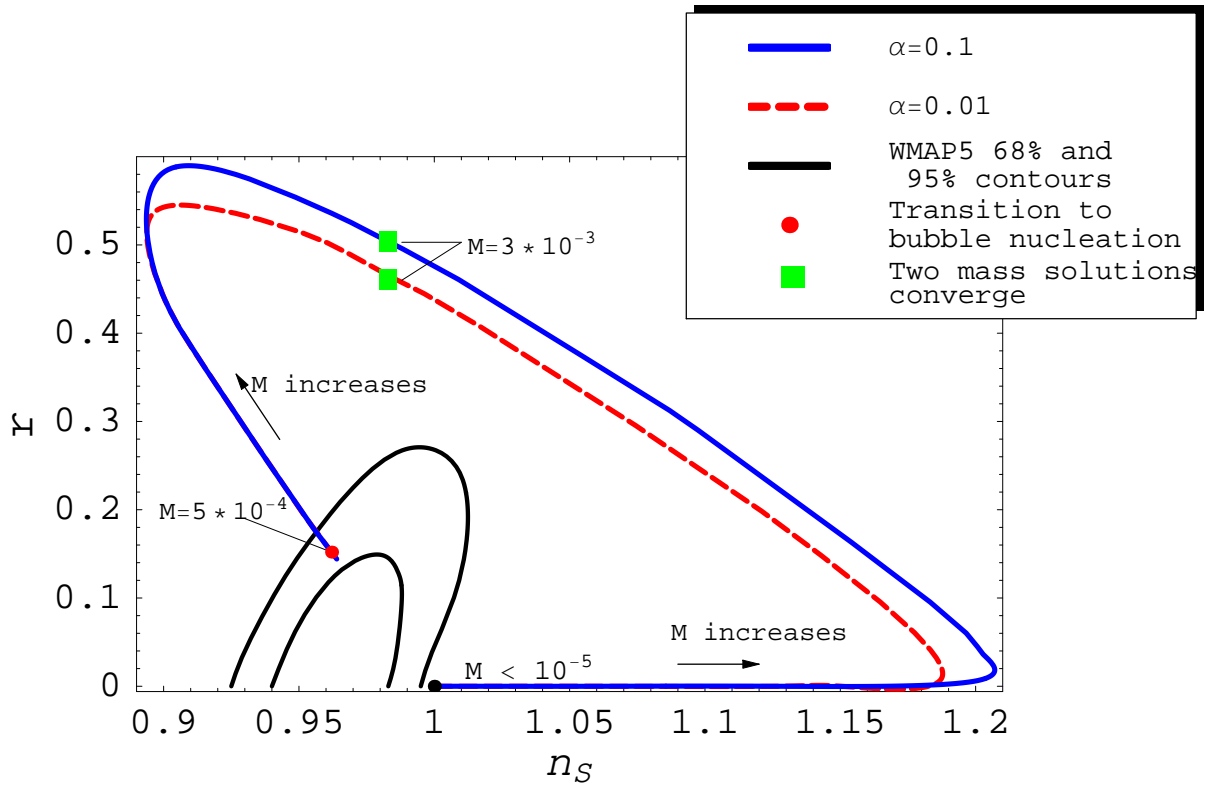}
\caption{The trajectories described in the $n_{\rm S}-r$ parameter
plane for first-order models when $M$ is varying and $m$ is set by the
CMB normalization. The two lines correspond to different values of the
coupling constant $\alpha$, the outermost $\alpha=0.1$
and the innermost $\alpha=0.01$. The two endpoints correspond to
the endpoints in Fig.~\ref{2m} and converge at $M \sim 10^{-3} m_{\rm
Pl}$ corresponding to the union of the branches in Fig.~\ref{2m}, at
$(n_{\rm S}, r) \sim (0.99,0.5)$. Mass values are given in Planck
units.}\label{nr1}
\end{figure*}

To determine $\phi_{\rm crit}$ we need to find the value at which the
percolation parameter reaches unity, $p_{\rm crit} \sim 1$. Solving 
Eq.~(\ref{perc}), we get
\begin{equation}
S_{\rm crit}\sim  \ln \frac{\lambda M^4}{4 p_{\rm crit} H^4}
\label{S_cr} \,,
\end{equation}
where $S_{\rm crit}$ is given by Eq.~(\ref{4act}).

Inverting Eq.~(\ref{S_cr}) yields a value for $\phi_{\rm crit}$ (only
one of the three roots lies in the allowed range) and in turn this
allows us to determine $\phi_{\rm end}$, and, by comparison with
$\phi_{\rm \epsilon}$, the mechanism by which inflation ends.

Knowing $\phi_{\rm end}$ we can calculate the value of the field at
horizon exit, $\phi_{\rm 55}$. In this model $\phi$ rolls towards its minimum at $\phi=0$ so  $\phi_{\rm 55} > \phi_{\rm end}$. Using the expression for the number of $e$-folds between two field values $\phi_1$ and $\phi_2$ we get,
\begin{equation}
N(\phi_1,\phi_2)\equiv \ln \frac{a_2}{a_1}
\sim -\frac{8 \pi}{m_{\rm Pl}^2}
\int_{\phi_1}^{\phi_2}\frac{V}{V'} \, d\phi \,.
\end{equation}
For $\phi_1= \phi_{55}$ and  $\phi_2= \phi_{\rm end}$, and substituting for $V$, we have
\begin{equation}\label{Ne}
N(\phi_{55},\phi_{\rm end})= 2 \pi \lambda \frac{M^4}{m^2 m_{\rm Pl}^2}
\ln \frac{\phi_{55}}{\phi_{\rm end}} + \frac{2\pi}{m_{\rm
    Pl}^2}(\phi_{55} ^2 - \phi_{\rm end}^2) \,,
\end{equation}
where we make no assumptions on the relative size of the
two masses and retain both terms. Substitution of $\phi_{\rm end}$ yields $\phi_{\rm 55}$ and now we can calculate the scalar spectral index, $n_{\rm S}$, and the tensor-to-scalar ratio, $r$, at horizon exit, by use of their expressions in terms of the usual slow-roll parameters,
\begin{eqnarray}\label{nS}
n_{\rm S}-1&=&-6 \epsilon + 2 \eta \,;\\
r&=& 16 \epsilon \,, \label{r}
\end{eqnarray}
where $\epsilon$ is given by Eq.~(\ref{eps}), and $\eta$ is
\begin{equation}\label{eta}
\eta \equiv \frac{m_{\rm Pl}^2}{8 \pi} \frac{V''}{V}
=\frac{m^2 \, m_{\rm Pl}^2}{2\pi (\lambda M^4 + 2 m^2 \phi^2)} \,,
\end{equation}
where the last equality is obtained by substitution of the
potential.

\begin{figure}[t]
\includegraphics[width=0.47 \textwidth]{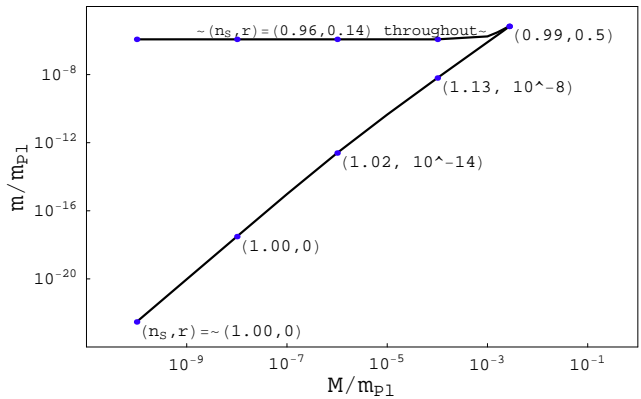}
\caption{The relation between the two mass scales. The WMAP
normalization admits two solutions for $m$, corresponding to 
false vacuum domination over the inflaton (lower branch),
and the opposite regime, for large $m$, which is nearly independent of
$M$ (upper branch). The two regimes converge to common behaviour. We
consider all three regimes in the analysis and set $\alpha=0.1$, given
by the upper curve in Fig.~\ref{nr1}.}\label{2m}
\end{figure}

At this point we can locate the model in the $n_{\rm S}-r$ plane and
determine its position in relation to WMAP5 confidence limits
\cite{komatsu}.

\subsection{Choosing parameters}

Throughout we set the self-interaction and coupling constants,
$\lambda$ and $\lambda '$ respectively, equal to unity. We are then
left with two constants, $\alpha$ and $\gamma$, and requiring the
energy density of the true vacuum to be zero fixes one of these in
terms of the other. We will fix $\alpha$ in terms of $\gamma$ but the reverse option could just as well be taken.

The CMB amplitude normalization can be used to relate the two masses.
We use this to fix the mass of the light field $\phi$ and then we are
left with only two undetermined parameters: the energy of the false
vacuum, $M$, and the constant $\alpha$. For each value of $\alpha$,
varying $M$ fully determines the dynamics of the fields, and describes
a trajectory in the $n_{\rm S}-r$ plane shown in Fig.~\ref{nr1}.

Each line is composed of two branches which correspond to the two
solutions of the WMAP normalization, and converge for large values of
$M \sim 2.7 \times 10^{-3} m_{\rm Pl}$. For values of $M$ larger than this there is no solution to the amplitude normalization hence no viable models. This can be seen also in Fig.~\ref{2m} which illustrates how the two different approximation schemes converge to a common behaviour and cease to exist after a certain value of $M$ (c.f.\ Fig.~1 of Ref.~\cite{copeland_al}). 

The right-hand branch in Fig.~\ref{nr1} corresponds to the lower
branch in Fig.~\ref{2m} and to the smaller value of $m$ from the WMAP
normalization. In this branch the approximate relation $M \sim
m^{2/5}$ (in Planck units) holds and the false vacuum dominates. The
dynamics are indistinguishable in the $n_{\rm S}-r$ plane when $M <
10^{-4} m_{\rm Pl}$. We start with the typical slightly blue tilted
spectrum and negligible tensor fraction. As $m$ continues to increase
so does the deviation from $n_{\rm S} \sim 1$ until the approximate
relation between the two masses breaks down and we have the inflaton
playing a more significant role in the relative contribution of the
two fields. At this point we observe a turn in the $n_{\rm S}-r$
plane, and the solution enters the intermediate region of comparable
field energy densities.

Despite this we still observe inflation ending by bubble nucleation
throughout this branch, from small values of $M$ to the maximum at $M
\sim 2.7 \times 10^{-3} m_{\rm Pl}$.

In the opposite branch, on the left-hand side, the model starts inside
the WMAP5 95\% confidence contour, well inside the inflaton dominated
regime. Similarly to the other branch we observe an initial period
where there is little dependence on the false vacuum energy,
corresponding to the plateau on Fig.~\ref{2m}, and the dynamics are
very well approximated by those of standard single-field inflation
with a $\phi^2$ potential, well known to satisfy WMAP5 data.

\begin{figure*}[t]
\includegraphics[width=0.8\textwidth]{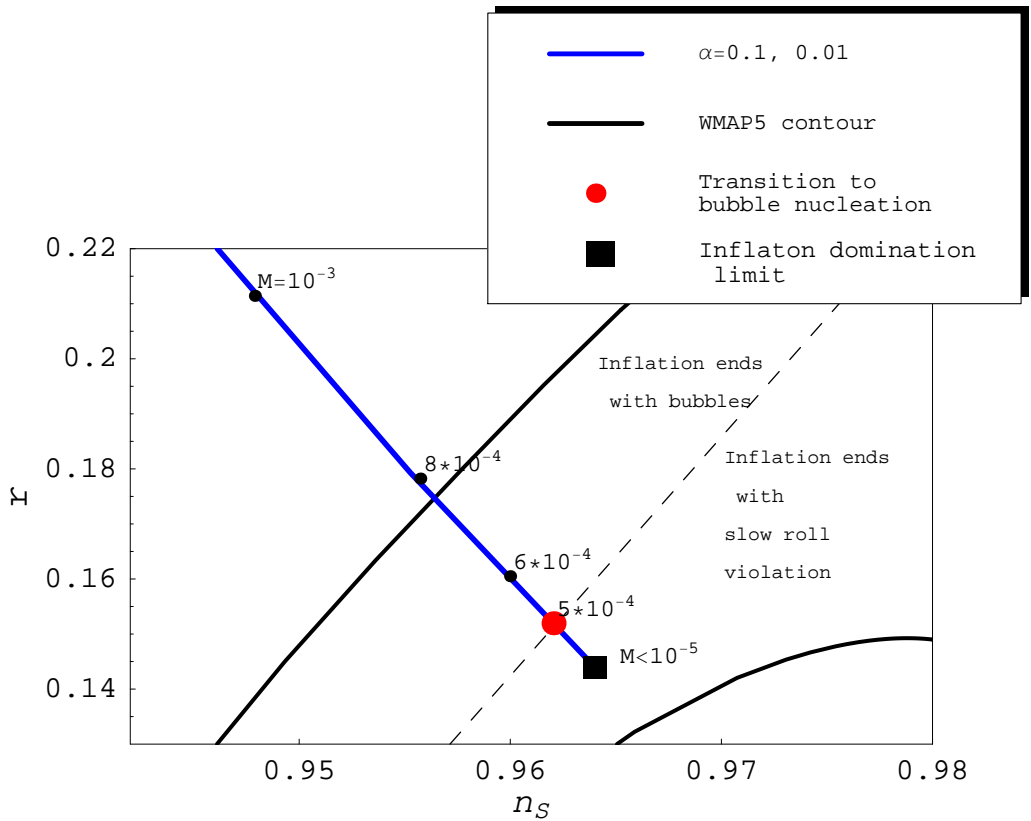}
\caption{Zoom of Fig.~\ref{nr1} showing the transition between models
ending by slow roll violation and bubble inflation. The transition
happens around $M \sim 5\times 10^{-4} m_{\rm Pl}$, well within the
WMAP5 95\% confidence contour. Mass values in Planck mass
units.}\label{zoom}
\end{figure*}

This regime breaks down as the false vacuum energy increases and
eventually we recover the regime where the phase transition triggers
the end of inflation before the violation of slow roll, meaning we are
again in the bubble production scenario.  The interesting results here
draw from the fact that the transition occurs inside the WMAP5 95\%
confidence contour, making these viable models even away from false
vacuum domination.  Fig.~\ref{zoom} is a zoom of this region showing
the field mass, $M$, at which the transition to bubble nucleation
occurs, $M \sim 5\times 10^{-4} m_{\rm Pl}$, still allowed by the 95\%
confidence limits.

\section{Three constraints}

In the previous section we looked at constraints in the $n_{\rm S}-r$
plane. By specifying a value for $\alpha$, one of our two free
parameters $(M,\alpha)$, the CMB normalization then allows us to
recover a trajectory in this plane and assess where the density
perturbations are compatible with WMAP5 data. We now compute other
constraints on the scenario, in the $M-\alpha$ plane.

\subsection{Model consistency}

We begin with the requirement that $M$ be not larger than an upper
limit above which, for a particular choice of couplings, the
transition does not complete ($\phi_{\rm crit}$ does not exist).  We
call this the model consistency constraint, which translates to a
relation for the value of $\phi_{\rm crit}$, coming from the
requirement that there exists a solution of Eq.~(\ref{delta}) for
$\delta$. Because of the constant term in Eq.~(\ref{delta})
this is an additional requirement to $0<\delta<2$.

Since we have chosen to set $m$ by the CMB normalization this can be
translated into an excluded region in the $(M,\alpha)$ plane (although
alternatively we could have expressed it in terms of a region in
$(M,m)$, by having $\alpha$ specified by the CMB normalization
instead). This yields the region below the upper (blue) curve in
Fig.~\ref{3constr}.  We see that specifying a value for the false
vacuum density imposes an upper limit on the coupling $\alpha$
(alternatively on the inflaton mass, $m$) in order for the model to
have the possibility to complete the phase transition.

\subsection{Big bubble constraint}

\begin{figure*}[t]
\includegraphics[width=0.8\textwidth]{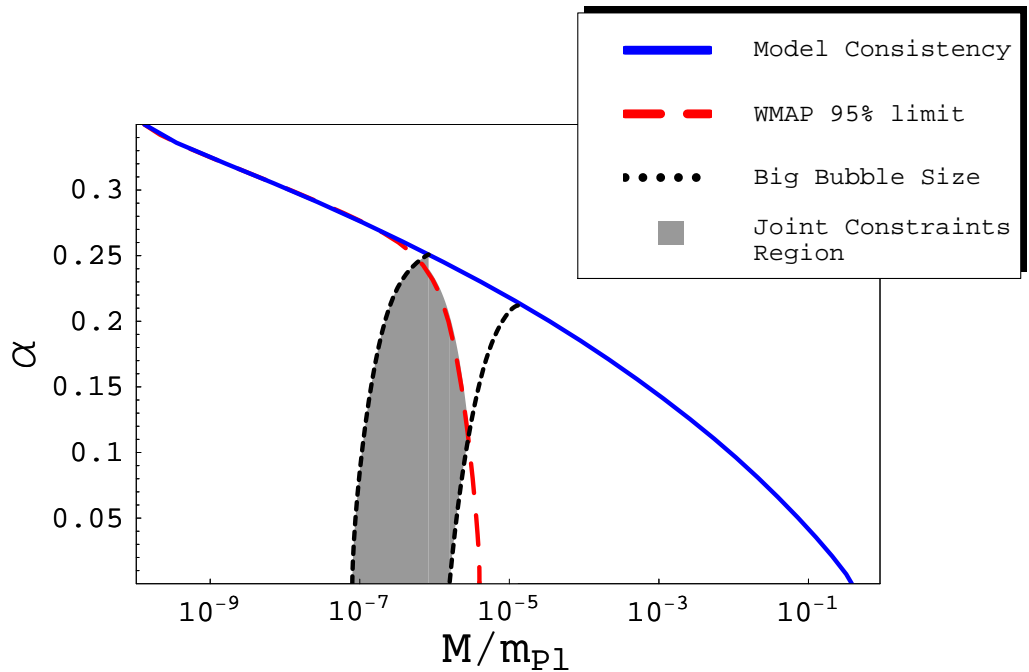}
\caption{Excluded regions in the $(M,\alpha)$ parameter space. The
continuous (blue) line corresponds to ensuring model consistency; if the
model lies above this region the phase transition will not take
place. The long dashed (red) line corresponds to WMAP5 constraints on the
value of the scalar perturbations tilt and allows models to the left of
the bound. The region between the short dashed (black) lines indicates
models satisfying the maximum size of bubbles allowed by the level
of anisotropy in the CMB.}
\label{3constr}
\end{figure*}

We adopt here a fairly crude criterion to judge whether the bubbles
are compatible with observations, which is that any bubbles produced
at the end of inflation and expanded to astrophysical sizes must,
during the epoch of recombination, have a comoving size not larger
than $20h^{-1}{\rm Mpc}$ \cite{liddle_wands91}. This corresponds to a
maximum filling fraction at that time of $10^{-5}$, and puts an upper
bound on the percolation rate of bubbles at the time the scales we
observe today left the horizon:
\begin{equation}
\left(\frac{\Gamma}{H^4}\right)_{55} \leq 10^{-5} \,.
\end{equation}
With our form for the action Eq.~(\ref{4act}) and choice of potential this
becomes 
\begin{equation}
S_{55}\sim -2.9 +4 \ln{\frac{m_{\rm Pl}}{\lambda^{1/4} M}}+ 11.5 \,.
\end{equation}
This gives us the region between the short dashed (black) lines in
Fig.~\ref{3constr}.

\subsection{WMAP constraint}

We can similarly place constraints on the $(M,\alpha)$ plane, by
considering the 95\% confidence limit resulting from the WMAP5 $n_{\rm
S}-r$ plane when tensors are included.
\begin{equation}
n_{\rm S} \lesssim 1.05 \,.
\label{nSwmap}
\end{equation}
Inverting Eq.~(\ref{nSwmap}) gives us an upper limit on $M$ in terms
of $\alpha$, resulting in the region left of the long dashed (red) line in
Fig.~\ref{3constr}.

We also see from Fig.~\ref{3constr} that this constraint is
opposed to that coming from the CMB maximum bubble size requirement, as
we argued in Section \ref{intro}. Big bubbles at last scattering put
an upper limit on the nucleation rate at horizon crossing while CMB
constraints on the spectral tilt put a lower bound on the nucleation
rate, from the requirement that $n_{\rm S}$ is not too distant from
scale invariance.

Nevertheless, a region of parameter space survives all constraints.

\section{Conclusions}

Our principal conclusion is that there do remain Einstein gravity
models of first-order inflation which are compatible with
observations, despite the increasing tension between the need for a
scale-invariant primordial spectrum and the suppression of large-scale
bubbles. We have exhibited a particular class of model and found the
parameter region where the first-order model is viable. Its
predictions for $n_{\rm S}$ and $r$ are similar to the simple $m^2
\phi^2$ slow-roll inflation model, though a little further from
scale-invariance. 

In this paper we have imposed a relatively simple constraint on the
bubbles, and have then assumed that their impact on the CMB is
negligible as far as constraints on the primordial perturbations are
concerned. A more detailed treatment would combine the two
perturbation sources and refit to the CMB data, which may lead to some
modification to the outcome in regimes where the bubble production is
close to the observational limit. For models where the
bubbles are safely within the observational limits this is not an
issue. 

This paper demonstrates that we are still some way from having a clear
view as to how the inflationary period of the Universe may have
ended. The literature contains three different mechanisms ---
violation of slow-roll, a second-order instability during slow-roll,
and bubble nucleation --- and we have shown that the last (and least
popular) of these remains a viable option. First-order models are of
phenomenological interest as the bubble spectrum is an additional
source of inhomogeneity that could be considered in matching
high-precision observations. The bubble collisions may also generate
detectable gravitational waves \cite{hogan, kosowsky_al,
huber_konstandin, caprini_al}.  There is therefore an ongoing need to
refine understanding of the nature of perturbations induced by a
primordial bubble spectrum.


\begin{acknowledgments}
M.C.\ was supported by FCT (Portugal) and by the Director, Office of
Science, Office of High Energy Physics, of the U.S. Department of
Energy under Contract No.\ DE-AC02-05CH11231.  A.R.L.\ was supported
by STFC (UK). We thank Andy Albrecht, Katie Freese, Andrei Linde, and
Eric Linder for discussions 
and comments.
\end{acknowledgments}

\begin{figure*}[t]
\includegraphics[width=0.8\textwidth]{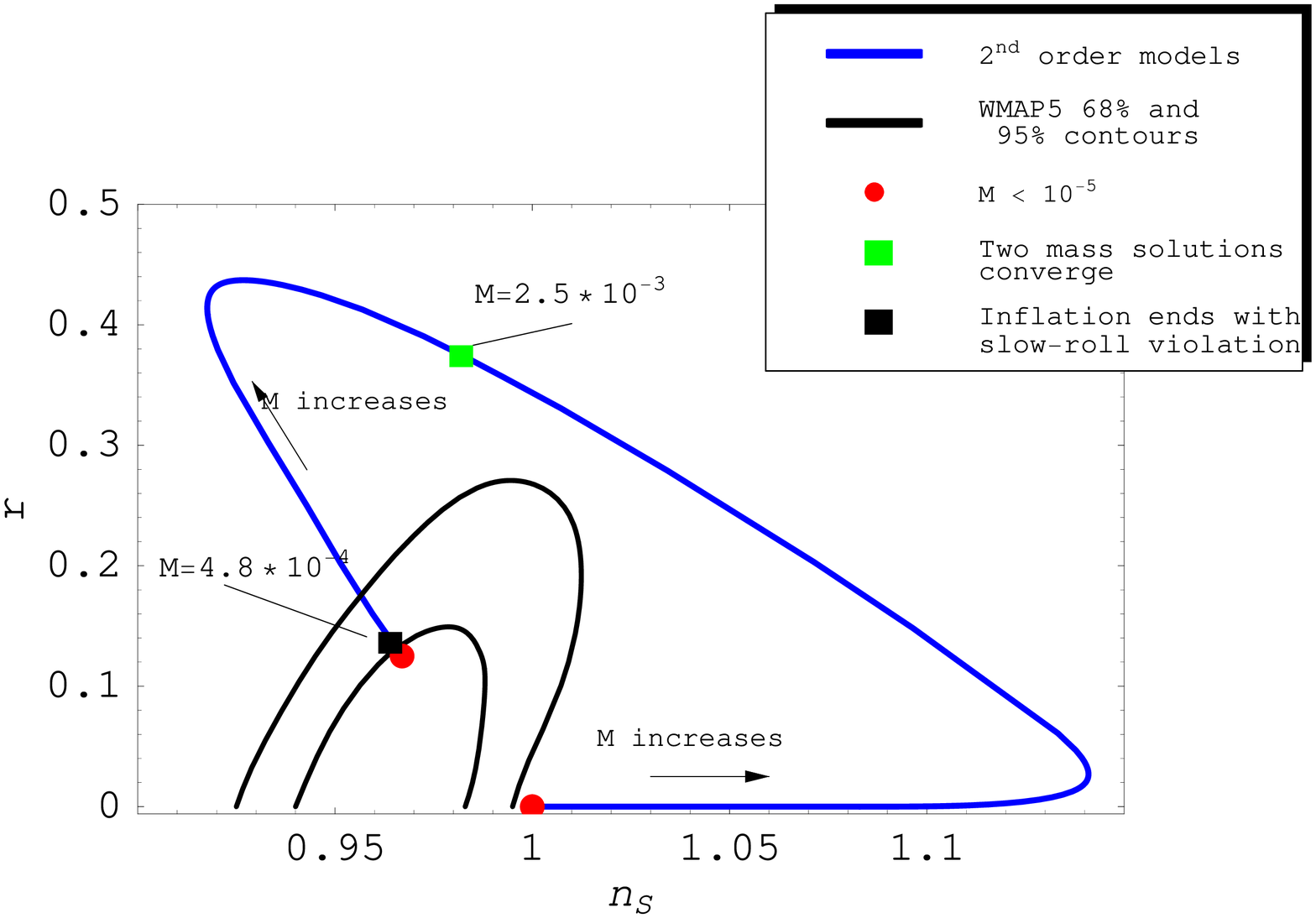}
\caption{Trajectory in parameter space $n_{\rm S}-r$ describing
  second-order hybrid inflation models when $M$ evolves from small
  through large values. The potential has one less coupling compared
  to the first-order case and all models are described by a single
  curve, as opposed to Fig.~\ref{nr1}. Allowed models are those which
  approximate slow roll behaviour. False vacuum, blue tilted, models
  all lie outside the 95\% C.L.}\label{nr2}
\end{figure*}

\appendix
\section{The second-order model}
\label{sec2nd}

Although not part of our main study, the full parameter range of the
second-order hybrid inflation model \cite{linde_2nd_I, linde_2nd_II,
copeland_al} is easily studied using the machinery we have used for
the first-order case.  The second-order model also uses the energy
density of an auxiliary field to raise the energy scale for inflation
without endangering slow roll. The phase transition in this case is
continuous, with the $\psi$ field rolling down to the true vacuum (see
Fig.~\ref{2pot}(a)). There are no bubbles now and hence no bubble
constraint; we just have to consider whether the usual perturbations
are compatible with WMAP5 data.  Furthermore since now there is no
cubic term to break the degeneracy between the two minima, there is
the possibility of topological defect formation at the end of
inflation, as different regions in space roll towards one or the other
minimum.  However we do not consider their possible impact here.

The dynamics are closely related to those in the first-order case. The
critical point where the phase transition completes is a point of
instability $\phi_{\rm inst}$, after which $\psi=0$ becomes unstable
and starts to roll.  The potential for this case is a
particularization of the first-order potential Eq.~(\ref{pot}) with
$\lambda=-\alpha$ and $\gamma=0$, and becomes,
\begin{equation}\label{pot2}
V(\phi,\psi)= \frac{1}{4} \lambda(\psi^2-M^2)^2+ \frac{1}{2} m^2
\phi^2 + \frac{1}{2} \lambda' \phi^2 \psi^2 \,.
\end{equation}

Apart from the expression determining $\phi_{\rm inst}$, we can retain
most of the expressions from the first-order model and build a similar
picture in the $n_{\rm S}-r$ plane. We present this in Fig.~\ref{nr2},
again for $\lambda = \lambda'=1$. We see that the false vacuum
dominated regime, which has $n_{\rm S}>1$ and negligible $r$, lies
entirely outside the WMAP5 allowed region, as does the main curve of
the intermediate regime. Only once the trajectory heads towards the
slow-roll limit does it become compatible with observations. At
$M\sim9 \times10^{-4} m_{\rm Pl}$ the models cross the WMAP5 95\%
contour and at $M\sim5 \times10^{-4} m_{\rm Pl}$ inflation ends
through slow roll violation instead of a phase transition.


\bibliographystyle{apsrev}
\bibliography{firstorder}


\end{document}